\documentclass[epj]{svjour} 

\usepackage{graphicx} \graphicspath{ {./figures/}{./figures_py/} }
\usepackage{amsmath} 
\usepackage{amssymb, bbm} 
\usepackage{amstext} 
\usepackage{booktabs} \cmidrulekern=.3em 
\usepackage{array}
\usepackage[utf8]{inputenc} 
\usepackage{placeins} 

\renewcommand\Re{\operatorname{Re}}

\newcommand\numberthis{\addtocounter{equation}{1}\tag{\theequation}}
\newlength{\feynwidth} 

\begin{document}

\title{Hyperon-nucleon three-body forces and strangeness in neutron stars}

\author{
Dominik Gerstung\inst{} \and
Norbert Kaiser\inst{} \and
Wolfram Weise\inst{}
}


\institute{
Physics Department, Technical University of Munich, 85748 Garching, Germany
}

\date{} 

\abstract{
Three-body forces acting on a $\Lambda$ hyperon in a nuclear medium are investigated, with special focus on the so-called hyperon puzzle in neutron stars. The hyperon-nucleon two-body interaction deduced from SU(3) chiral effective field theory is employed at next-to-leading order. Hyperon-nucleon three-body forces are approximated using saturation by decuplet baryons and are transcribed to density-dependent effective two-body interactions. These together are taken as input in a Brueckner-Bethe-Goldstone equation with explicit treatment of the $\Lambda N\leftrightarrow\Sigma N$ and $\Lambda NN\leftrightarrow\Sigma NN$ coupled channels. Single-particle potentials of a $\Lambda$ hyperon in symmetric nuclear matter and neutron matter are calculated. With parameters of the $\Lambda NN$ three-body force constrained by hypernuclear phenomenology, extrapolations to high baryon density are performed. By comparison of the $\Lambda$ and neutron chemical potentials at densities characteristic of the core of neutron stars it is found that the combined repulsive effects of two- and three-body correlations can make the appearance of $\Lambda$ hyperons in neutron stars energetically unfavourable, thus potentially offering a possible answer to a longstanding query. 
\PACS{
      {PACS-key}{describing text of that key}    }
}

\maketitle

\section{Introduction}
\label{sec:intro}

The existence of heavy neutron stars with masses around 2$M_\odot$ \cite{Demorest2010,Antoniadis2013,Fonseca2016,Cromartie2019} sets strong constraints on the equation of state (EoS) of dense baryonic matter. This EoS must be sufficiently stiff, i.e. the pressure $P(\cal{E})$ at energy densities $\cal{E}\sim$ 1 GeV/fm$^3$ must be large enough to support such massive compact objects against gravitational collapse. The detection of gravitational wave signals from two merging neutron stars \cite{Abbott2017}  adds further important information on the EoS, by providing limits for the tidal deformability and for neutron star radii \cite{Most2018,De2018}. 

The composition and properties of strongly interacting matter at high baryon densities is a topic of continuing interest. Various options are under discussion. A time-honored description of matter in the core of neutron stars uses hadronic degrees of freedom (baryons and mesons) with strong many-body correlations \cite{APR1998}. A modern version that arrives at a similar EoS, consistent with observations, is based on a chiral nucleon-meson field theory combined with functional renormalization group methods \cite{DW2015,DW2017}. In this latter approach neutron star matter can be represented as a relativistic Fermi liquid in the sense of Landau theory \cite{FW2019}. The density-dependent leading Landau parameters show the characteristic behaviour of a strongly correlated fermionic many-body system, albeit less extreme in comparison with another well-known Fermi system at low temperature, namely liquid $^3$He. Alternative descriptions of neutron star matter involve a (possibly smooth) transition from hadronic matter to some form of quark matter \cite{Baym2018,Baym2019,McLerran2019,FFM2019}. 

The required stiffness of the EoS implies strong restrictions on the appearance of hyperons in neutron star matter and for the underlying hyperon-nuclear interactions. The role of hyperons in neutron star matter has been the subject of a wide range of investigations, with varying conclusions, during the past two decades, \cite{Takatsuka2002,Nishizaki2002,Takatsuka2008,Stone2007,Stone2010,Djapo2010,Vidana2011,Yamamoto2013,Hell2014,YFYR2014,BS2015,Lonardoni2015,Vidana2019}, mentioning also the pioneering work of Ref.~\cite{ambart} that appeared six decades ago. A naive introduction of $\Lambda$ hyperons, with only two-body $\Lambda N$ interactions, would appear to be energetically favorable by replacing neutrons at baryon densities around 2-3 $\rho_0$ (in terms of the equilibrium density of normal nuclear matter, $\rho_0 = 0.16$ fm$^{-3}$). However, then the EoS of neutron star matter would become far too soft and unable to satisfy the two-solar-mass constraint \cite{Djapo2010,Hell2014}. As a possible option to deal with this problem, the introduction of a strongly repulsive hyperon-nuclear three-body force could prohibit altogether the appearance of $\Lambda$ hyperons in neutron stars \cite{Lonardoni2015}. A phenomenological analysis \cite{BS2015}, taking into account constraints from both hypernuclear physics and neutron star observations, points out indeed that the interaction between $\Lambda$ hyperons and dense matter has to become repulsive already at densities below three times $\rho_0$. 

The aim of the present work is to investigate whether and to what extent a microscopic description of hyperon-nucleon two- and three-body forces is capable of providing the necessary repulsion in dense baryonic matter. The key quantities to be calculated are the hyperon ($\Lambda$ and $\Sigma$) single-particle potentials $U_{\Lambda,\Sigma}(\rho)$ in nuclear and neutron matter. The density dependence of the resulting chemical potential $\mu_\Lambda(\rho)= M_\Lambda +U_{\Lambda}(\rho)$  of a $\Lambda$ in neutron matter indicates whether it is favourable to replace neutrons by $\Lambda$ hyperons in the core of neutron stars. 

Our starting point is the hyperon-nucleon interaction derived from SU(3) chiral effective field theory (ChEFT) at next-to-leading order (NLO) \cite{Haidenbauer2013}. This interaction includes exchanges of one and two pseudoscalar octet mesons and four-baryon contact terms with SU(3)-symmetric low-energy constants fitted to the (admittedly scarce) hyperon-nucleon scattering data.  Two versions of this interaction exist and will both be used in the present work: NLO13 \cite{Haidenbauer2013}, and a more recent edition, NLO19 \cite{Haidenbauer2019}. Three-body forces
enter first at next-to-next-to-leading order (NNLO) in this scheme \cite{Petschauer2016}. Estimating the strength of contact terms in these forces through the contributions of explicit baryon decuplet resonances in intermediate states, these three-body interactions are promoted from NNLO to NLO \cite{Petschauer2017} due to the small decuplet-octet baryon mass-splitting. We shall adopt here the same strategy. 

Previous calculations of hyperon-nuclear potentials using Brueckner theory with NLO13 interactions as input have been reported in refs.\,\cite{Petschauer2016b,Haidenbauer2017,Kohno2018}. The present study builds on these results, with extensions in several directions. Special attention will be paid to the importance of the $\Lambda N\leftrightarrow\Sigma N$ transition potential, a necessary ingredient of any such calculation. In particular, the $\Lambda NN\leftrightarrow\Sigma NN$ coupled channels in the three-body sector, translated into density-dependent effective two-body potentials as in ref.\,\cite{Petschauer2017}, will be treated here for the first time explicitly when solving coupled-channel Bethe-Goldstone equations.

\section{Basic baryonic interactions}
\label{sec:int}

\subsection{Hyperon-nucleon two-body interactions}

\begin{figure}[t]
\centering
\includegraphics[width=\feynwidth]{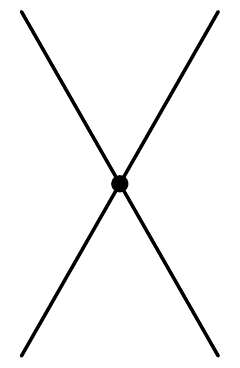} \quad
\includegraphics[width=\feynwidth]{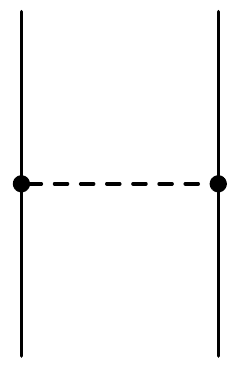}\\[.4\baselineskip]
\includegraphics[width=\feynwidth]{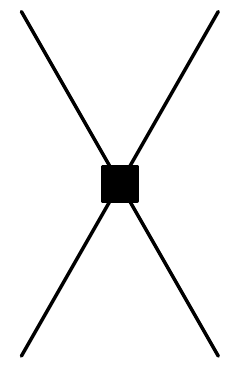} \quad
\includegraphics[width=\feynwidth]{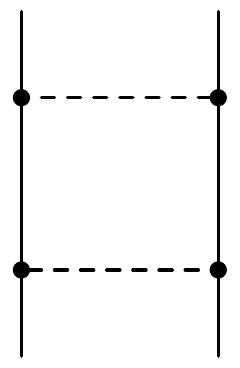}\quad
\includegraphics[width=\feynwidth]{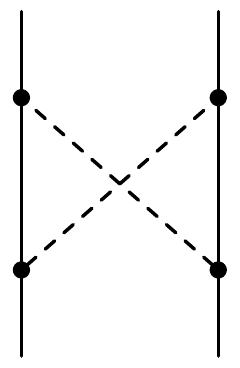}\quad
\includegraphics[width=\feynwidth]{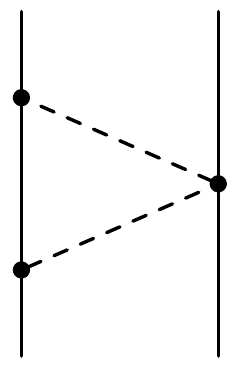}\quad
\includegraphics[width=\feynwidth]{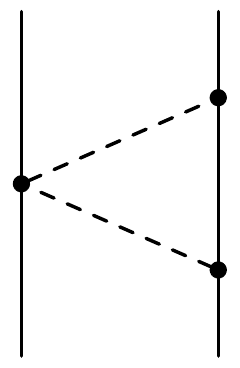}\quad
\includegraphics[width=\feynwidth]{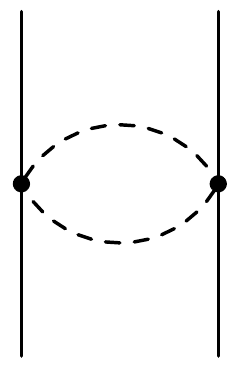}
\caption{Leading and next-to-leading order diagrams representing the baryon-baryon interaction potential. Solid and dashed lines denote octet baryons (\(N,\Lambda,\Sigma\)) and mesons (\(\pi,K,\eta\)), respectively.}
\label{fig:pwrcounting}
\end{figure}

For the description of the hyperon-nucleon two-body interaction, SU(3) chiral effective field theory is used up to NLO with the Weinberg power counting applied to the potential, as reported in detail in ref.\,\cite{Haidenbauer2013}. The leading order potential involves the exchange of a single pseudoscalar octet meson $(\pi, K, \eta)$ and non-derivative four-baryon contact terms.  At NLO two-meson exchange diagrams at one-loop level arise together with  additional contact terms with explicit momentum dependence of order $p^2$ (see fig.\,\ref{fig:pwrcounting}). 
The set of SU(3)-symmetric contact terms represents unresolved short-distance dynamics. The corresponding low-energy constants (LECs) are fitted to low-energy hyperon-nucleon scattering data and the hypertriton \({}^3_\Lambda H\) binding energy (and partly to elastic $NN$-scattering phase shifts). SU(3) symmetry breaking effects are incorporated through the physical masses of the baryons and of the exchanged pseudoscalar mesons. Further details on input parameters of the two versions, NLO13 and NLO19, can be found in refs.\,\cite{Haidenbauer2013,Haidenbauer2019}. 

 Solving the coupled-channel Lippmann-Schwinger equations using these chiral potentials involves a regulator function of the form \(\exp[-(p'^4+p^4)/\lambda^4]\) depending on the momenta, $p$ and $p'$, of the in- and outgoing baryons. The cutoff $\lambda$ is chosen in the range 500 ... 600 MeV. 

 A characteristic feature of the resulting $\Lambda N$ S-wave phase shifts that distinguishes the chiral EFT-based potential from all earlier phenomenological interactions, such as J\"ulich'04 \cite{HM2005} and Nijmegen NSC97f \cite{Rijken1999}, is the following. As there is no (leading-order) one-pion exchange $\Lambda N$ interaction term, the driving intermediate range attraction in this channel comes from the iterative mechanism $\Lambda N \rightarrow \Sigma N \rightarrow \Lambda N$, involving the exchange of two pions. The $\Lambda\Sigma\pi$ coupling as given by SU(3) symmetry implies strong attraction through this second-order process
that is partly balanced by the strong short-distance repulsion. The net attraction is seen in the low-energy $^3S_1$ $\Lambda N$ phase shift. The $^1S_0$ phase shift also shows this attractive behavior at low energies but changes sign around a lab momentum of 600 MeV and becomes repulsive. This momentum-dependent repulsion is generated by the short-distance terms at NLO, as can be clearly seen by selectively turning off the $\Lambda\Sigma\pi$ coupling \cite{Haidenbauer2017}. The repulsive effects in the $\Lambda N$ interaction at high momenta have their direct impact on the $\Lambda$ single-particle potentials in nuclear and neutron matter which change from attraction to repulsion at densities around or below $2\,\rho_0$, a feature not matched by any of the phenomenological potentials. However, as we shall see, such two-body repulsion is still not strong enough to prevent $\Lambda$ hyperons from occuring at baryon densities encountered in neutron stars and thereby softening the EoS to an unwanted degree.

The NLO13 and NLO19 interactions give almost identical results in their comparison with the available hyperon-nucleon scattering cross sections and $S$-wave phase shifts. NLO19 features a slightly weaker $\Sigma N$ coupling than NLO13, with correspondingly readjusted contact terms. Although these differences are of no significance in reproducing the empirical two-body YN data, they show up in the predicted density dependence of the $\Lambda$-nuclear single-particle potential. Computations of this potential with both NLO13 and NLO19 interactions are therefore instructive as they give an impression of possible uncertainties in extrapolations to higher baryon densities. In practice we use a default cutoff $\lambda = 500$ MeV for both interactions.

The unavoidable truncation in the hierarchy of hyperon-nucleon interactions at NLO is a possible source of uncertainties. It would certainly be desirable in the future to reach a comparable level of accuracy in YN interactions as it has been achieved for the chiral nucleon-nucleon interaction. However, at this point the limited hyperon-nucleon data base does not (yet) permit a meaningful systematic extension to NNLO.

\subsection{Three-body interactions of hyperons with nucleons}

Three-nucleon forces are an important element in the quantitative understanding of nuclear few- and many-body systems. Likewise, the $\Lambda NN$ three-body interaction is expected to play a significant role in hypernuclear systems. Our interest in the present work is directed, in particular, to the behavior of $\Lambda NN$ and $\Sigma NN$ three-body forces (3BF) in dense baryonic matter, where their influence is expected to grow continuously with increasing density.

\begin{figure}
\centering
\includegraphics[scale=0.5]{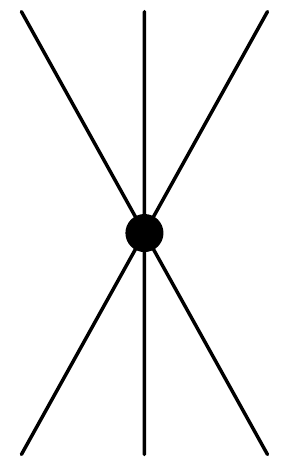}\qquad
\includegraphics[scale=0.5]{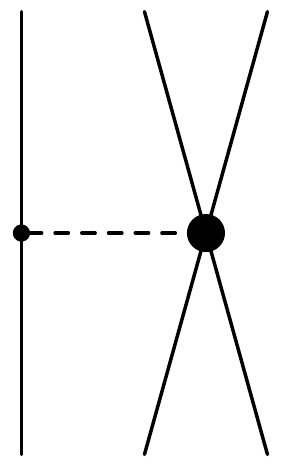}\qquad
\includegraphics[scale=0.5]{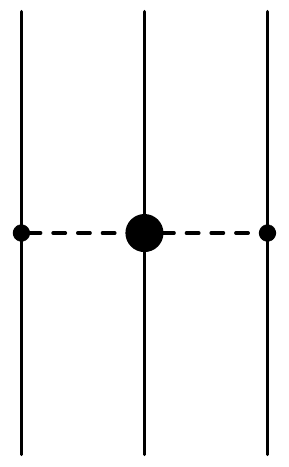}
\caption{
Leading three-baryon interactions: contact term, one-meson exchange and two-meson exchange.
\label{fig:3BF}
}
\end{figure}

In the systematic expansion of SU(3) chiral EFT, baryonic three-body forces appear at NNLO. Their detailed derivation is given in ref.\,\cite{Petschauer2016} on which we shall build. The leading 3BF diagrams are shown in fig.\,\ref{fig:3BF}. They fall into three classes: six-baryon contact terms, one-meson exchange pieces, and two-meson exchange pieces. The corresponding potentials are denoted by $V^{(0)}, V^{(1)}$ and $V^{(2)}$, respectively. The contact potential $V^{(0)}$ involves all possible combinations of baryon spin operators $\vec \sigma_1,  \vec \sigma_2, \vec \sigma_3$. The one-meson exchange 3BF potential has the following generic form:
\begin{equation}
V^{(1)}_i = \frac{1}{2f^2} \frac{\vec\sigma_1\cdot\vec q}{\vec q^{\,2}+m_i^2} (
A \vec\sigma_3+ B \, \mathrm i \vec\sigma_2\times\vec\sigma_3)\cdot\vec q
\,,
\end{equation}
where the inverse squared  pseudoscalar meson decay constant $f^{-2}= 4.8\,$fm$^2$ determines the interaction strength. The index $i\in \{\pi^0,\pi^+,\pi^-,K^+,K^-,K^0,\bar K^0,\eta\} $ refers to the exchanged meson with mass $m_i$, carrying the momentum transfer $\vec q$. The two parameters $A$ and $B$ are combinations of low-energy constants.
 
On the other hand, the two-meson exchange potential (depicted in fig.\,\ref{fig:3BF}) has following generic form:
\begin{align*}
V^{(2)}_{ij} ={} & -\frac{1}{4f^4} \frac{\vec\sigma_1\cdot\vec q\ ~~\vec\sigma_3\cdot\vec q'}{(\vec q^{\,2}+m_i^2)(\vec q'^{\,2}+m_j^2)} \\
&\qquad\times\Big(A' + B'\vec q\cdot\vec q' +D'\, \mathrm i\vec\sigma_2\cdot(\vec q\times\vec q')\Big) \,. \numberthis
\end{align*}
The parameters $A',B'$ and $D'$ are combinations of low-energy constants determined in the meson-baryon subsector. The transferred momenta $\vec q$ and $\vec q'$ are carried by the mesons with masses $m_i$ and $m_j$, respectively. The complete three-body potential is constructed by summing the contributions from all distinguishable diagrams with all possible exchanged pseudoscalar mesons, including baryon exchange operations where needed.

As they stand, the 3BF terms at NNLO introduce a prohibitively large number of low-energy constants in the contact potentials $V^{(0)}$. The existing data-base is too limited and does not permit a meaningfully constrained determination of {all these parameters. However, their number can be reduced substantially by employing an approximate scheme referred to as decuplet saturation: contact vertices are resolved by propagating explicit decuplet baryons in intermediate states.  In the case of the $2\pi$-exchange $3N$-interaction, this approximation is motivated by the well-known fact that the $\Delta(1232)$ dominates $P$-wave pion-nucleon scattering and therefore enters prominently through the two-pion exchange mechanism \cite{Miyazawa1957} shown in Fig.\,\ref{fig:3BFdec} (diagram on the right). Likewise, chiral dynamics with explicit $\Delta$ degrees of freedom proved to be a successful starting point for approaching the nuclear many-body problem \cite{Fritsch2005}. The decuplet dominance approximation is a natural extension of these considerations to $SU(3)$ and hyperon-nuclear interactions. As already mentioned, it has the welcome feature of promoting three-baryon forces from NNLO to NLO within the chiral hierarchy. In essence, the NNLO diagrams in fig.\,\ref{fig:3BF} are replaced by the NLO diagrams of fig.\,\ref{fig:3BFdec}. In our subsequent application, two of the in- or outgoing baryons are nucleons (proton, neutron), while the third one is a strangeness $S=-1$ hyperon ($\Lambda$ or $\Sigma$) and only the (long-range) pion-exchange is considered. The decuplet intermediate states are then either $\Sigma^*(1385)$ or $\Delta(1232)$. The heavier pseudoscalar mesons (kaons and $\eta$ meson) contribute at shorter distances comparable to the inverse cutoff $\lambda^{-1}\sim 0.4$ fm. Their effects are understood to be included in the parameters of the pertinent contact terms.

\begin{figure}
\centering 
\includegraphics[scale=0.5]{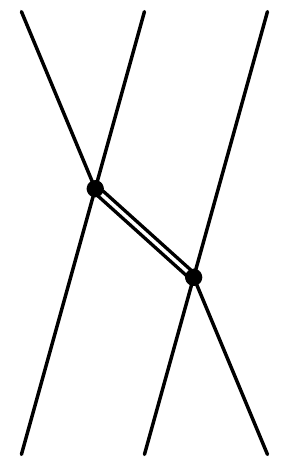}\qquad
\includegraphics[scale=0.5]{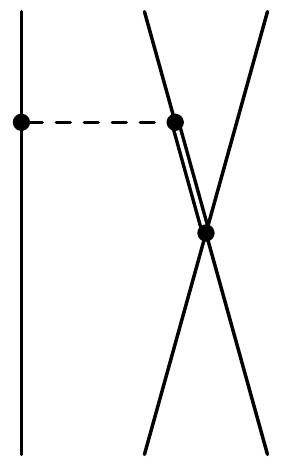}\qquad
\includegraphics[scale=0.5]{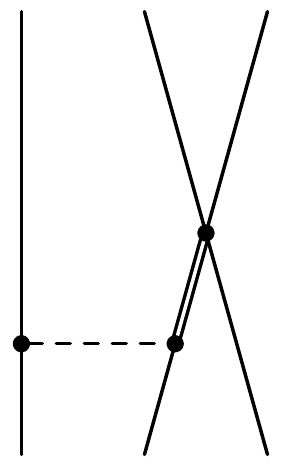}\qquad
\includegraphics[scale=0.5]{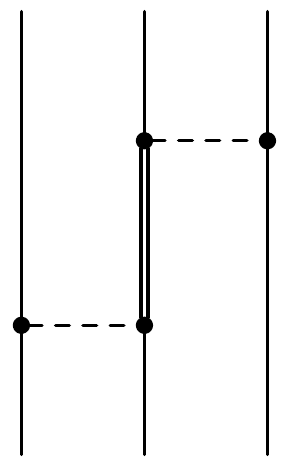}
\caption{Three-baryon forces with explicit decuplet baryons in the intermediate states (represented by double lines) \cite{Petschauer2017}.} 
\label{fig:3BFdec}
\end{figure}

In what follows we use the detailed formalism developed in refs.\,\cite{Petschauer2016,Petschauer2017}. Within the decuplet dominance approximation there are only three remaining constants to be determined: the coupling strength $C$ of the transition vertex between an octet baryon and a decuplet baryon with a pseudoscalar meson absorbed or emitted; and two coupling constants, $H_1$ and $H_2$, for the four-point vertices connecting three octet baryon lines and one decuplet baryon \cite{Petschauer2016,Petschauer2017}. The coupling $C$ is uniquely determined by the $\pi N\rightarrow\Delta$ transition vertex which is in turn constrained by the decay width $\Gamma(\Delta\rightarrow \pi N)\simeq 115$ MeV. The (large-$N_c$) value $C = 3g_A/4 \simeq 0.95$ with the nucleon axial vector coupling constant, $g_A = 1.26$, is well compatible with this constraint. The remaining two constants, $H_{1,2}$ of dimension $(length)^2$, are still free to choose. We use hypernuclear phenomenology in order to restrict their possible values.

Next, the three-body interactions in the $\Lambda NN\!\rightarrow\!\Lambda NN$, $\Lambda NN\leftrightarrow\Sigma NN$,  and $\Sigma NN \rightarrow\Sigma NN$ coupled channels are translated into effective density-dependent hyperon-nucleon potentials, which additionally enter the coupled Bethe-Goldstone equations.  This is done as in refs.\,\cite{Petschauer2016,Petschauer2017,Haidenbauer2017,Holt2010} by integrating one of the two nucleons over the filled Fermi sea,  as illustrated schematically in fig.\,\ref{fig:3BFeff}:
\begin{figure}
\centering 
\includegraphics[scale=0.2]{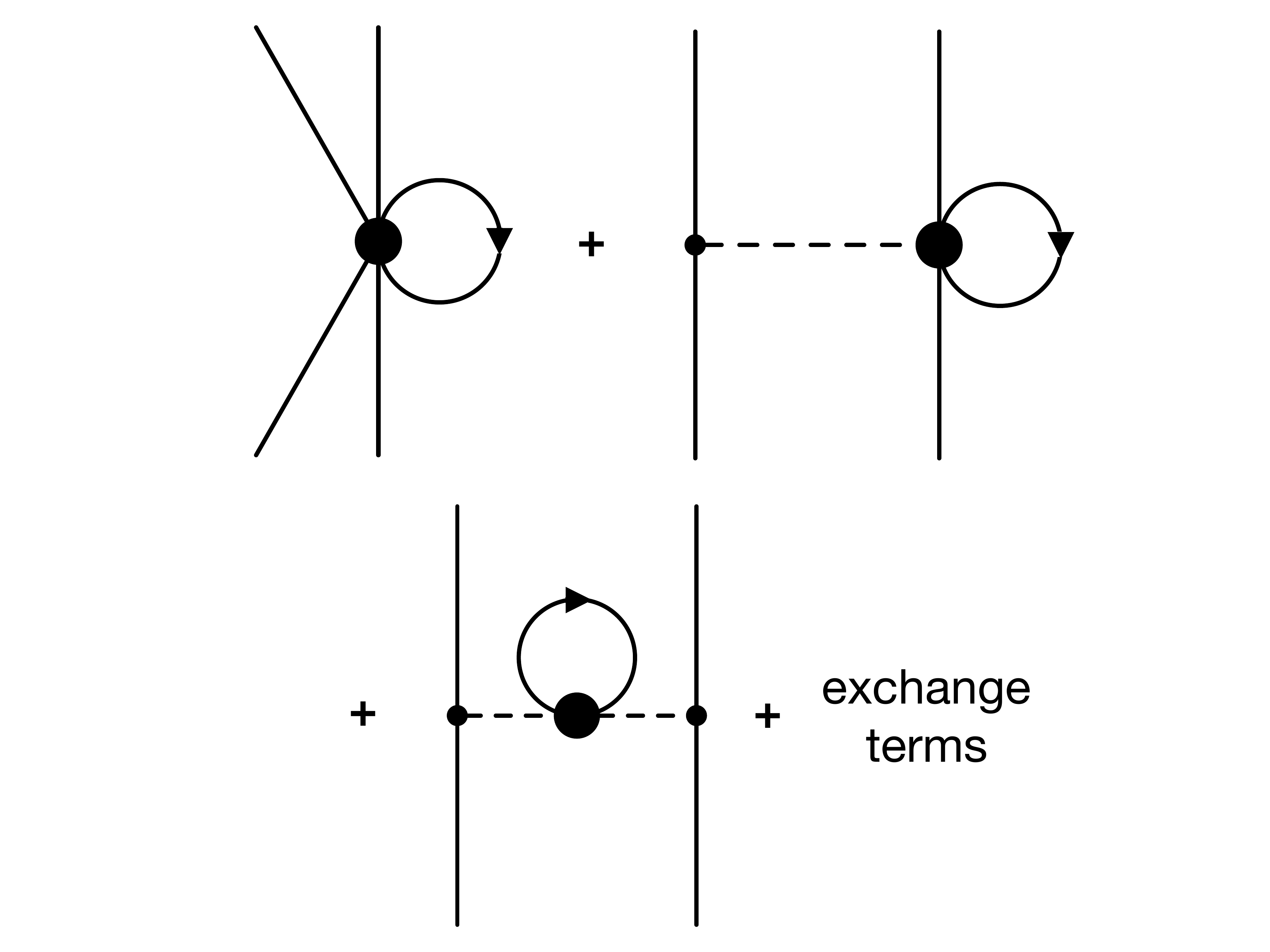}
\caption{Density-dependent in-medium potentials generated from 3-body forces. Upper left: contact term; upper right: one-meson exchange term; lower part: two-meson exchange term. Loops with arrows indicate integration over the filled nucleon Fermi sea. } 
\label{fig:3BFeff}
\end{figure}

\begin{equation} \label{eq:red}
V_{12}^{\text{eff}}(\rho_p,\rho_n) = \sum_{N=p,n}\int\limits_{|\vec k|\leq k_F^{(N)}}\frac{\mathrm d^3k}{(2\pi)^3}\,{\mathrm tr}_3V_{123} \,,
\end{equation}
where ${\mathrm tr}_3$ denotes the spin-trace over the third particle and the summation over nucleons (protons and neutrons) in the Fermi sea is performed. The resulting effective two-body potentials depend explicitly on the proton and neutron densities,
\begin{equation}
\rho_p = {(k_F^{(p)})^3\over 3\pi^2}~,~~~\rho_n = {(k_F^{(n)})^3\over 3\pi^2}~.
\end{equation}
As a first example, we present expressions for the  $\Lambda N\to \Lambda N$ effective two-body potential in (asymmetric) nuclear matter with density $\rho = \rho_p + \rho_n$. Only the $\Lambda n $ effective potential in the nuclear medium needs to be given since the $\Lambda p$ potential follows by simply interchanging $\rho_n\leftrightarrow\rho_p$. The three parts have the following structure \cite{Petschauer2017}:\\ 
Contact term:
\begin{equation}
V^{(0)}_{\text{eff}}(\Lambda n) = {(H_1+3H_2)^2\over 18 \Delta}\,(\rho_n+2\rho_p)~,
\end{equation} 
with the decuplet-octet baryon mass difference $\Delta$, for which we take an average value of $\Delta = 270$ MeV.\\
One-pion exchange part:
\begin{equation}
V^{(1)}_{\text{eff}}(\Lambda n) =
{g_A C\over 9f^2 \Delta}(H_1+3H_2)\left(\rho_n+2\rho_p - {m_\pi^2\over 2\pi^2}\Gamma_0\right),
\end{equation}
where $\Gamma_0(p;k_F^{(N)})$ is a function of the $\Lambda N$ center-of-mass momentum $p$, and the proton or neutron Fermi momentum $ k_F^{(N)}$, given explicitly in ref.\,\cite{Petschauer2017}.\\
Two-pion exchange part:
\begin{equation}
V^{(2)}_{\text{eff}}(\Lambda n) = {g_A^2 C^2\over 6f^4 \Delta}\left[\rho_n+2\rho_p+ {\cal F}(p,q;k_F^{(n)},k_F^{(p)})\right]\,,
\end{equation}
where  the function ${\cal F}$ depends additionally on the momentum transfer $q$ in the reduced two-body system. It is a lengthy expression involving the spin-orbit operator $\mathrm i \vec \sigma_2\cdot(\vec q\times \vec p\,)$, that is explicitly  given in eq.(46) of ref.\,\cite{Petschauer2017}.

Next, we present some selected expressions for the effective $\Lambda N \rightarrow \Sigma N$ interaction. As representative examples, consider the direct terms, shown in fig.\,\ref{fig:3BFeff}, of the $\Lambda n\rightarrow \Sigma^- p$ effective transition potential. Exchange terms are also included in the actual calculations but are not displayed here for simplicity. In the case of symmetric nuclear matter,  the corresponding lengthy formulas can be found in subsection III.D of ref.\,\cite{Petschauer2017}.\\
Contact term:
\begin{eqnarray} && V^{(0)}_\text{{eff}}(\Lambda n\!\rightarrow\!\Sigma^- p) = {1\over 6\sqrt{6}\Delta}\bigg\{(\rho_p\!+\!\rho_n) \big[4H_2^2\!-\!(H_1\!+\!H_2)^2\big]\nonumber\\ &&+ \vec\sigma_1\!\cdot\!\vec\sigma_2\bigg[\bigg(H_1^2\!+\!H_2^2\!+\!{10\over 3}H_1 H_2\bigg)\rho_n+ (H_1\!+\!3H_2)^2{ \rho_p\over 3}\bigg]\bigg\}\,. \nonumber\\ &&
\end{eqnarray} 
One-pion exchange part:
\begin{eqnarray}
&&V^{(1)}_\text{{eff}}(\Lambda n\!\rightarrow\!\Sigma^- p)={\sqrt{6}g_A C \over 27f^2\Delta}\, {\vec\sigma_1\cdot\vec q\,\vec\sigma_2\cdot\vec q\over m_\pi^2+\vec q^2}\nonumber\\ && \qquad\qquad \times \Big[(2H_1-3H_2)\rho_n-(2H_1+9H_2)\rho_p\Big]\,.
\end{eqnarray}
Two-pion exchange part:
\begin{eqnarray}
V^{(2)}_\text{{eff}}(\Lambda n\!\rightarrow\!\Sigma^- p)=-{16g_A D C^2\over 9\sqrt{6}f^4\Delta}{\vec\sigma_1\!\cdot\!\vec q\,\vec\sigma_2\!\cdot\!\vec q\over( m_\pi^2+\vec q^2)^2}(\rho_p+\rho_n) \vec q^2 , \nonumber\\ &&\end{eqnarray}
where the  SU(3) axial vector coupling constant $D$ stems from the $\Lambda\Sigma\pi$ vertex. For the $\Lambda p\rightarrow \Sigma^+ n$ channel, corresponding expressions hold with $\rho_n\leftrightarrow\rho_p$ interchanged.

Finally, the ladder summations in the coupled hyperon-nucleon channels require also as input the $\Sigma N\leftrightarrow\Sigma N$ effective two-body potentials in different charge combinations. As representative examples we list here expressions for the in-medium $\Sigma^-n\rightarrow\Sigma^- n$ potential:\\
Contact term:
\begin{eqnarray}
&&V^{(0)}_\text{{eff}}(\Sigma^- n) = 
 {1\over 6\Delta}\,\Big\{\big[(H_1+H_2)^2\rho_p + 4H_1^2\rho_n\big]
\nonumber\\ && \qquad \qquad\qquad\qquad -{1\over 3}\,\vec\sigma_1\cdot\vec\sigma_2\,(H_1+H_2)^2\,\rho_p\Big\}\, .
\end{eqnarray} 
One-pion exchange part:
\begin{eqnarray}
&&V^{(1)}_\text{{eff}}(\Sigma^- n)=
-{4g_A C\over 9f^2\Delta}\,{\vec\sigma_1\!\cdot\!\vec q\,\vec\sigma_2\!\cdot\!\vec q\over m_\pi^2+\vec q^2}\Big[H_1\rho_n \!+\!(H_1\!+\!H_2)\rho_p\Big]\,,\nonumber\\ &&
\end{eqnarray}
Two-pion exchange part:
\begin{equation}
V^{(2)}_\text{{eff}}(\Sigma^- n)=-{8g_A F C^2\over 9f^4\Delta}\,{\vec\sigma_1\cdot\vec q\,\vec\sigma_2\cdot\vec q\over( m_\pi^2+\vec q^2)^2}\,(\rho_n+\rho_p)\,\vec q^2\,,
\end{equation}
where now the SU(3) axial vector coupling constant $F$ enters. The $\Sigma^+ p\rightarrow\Sigma^+ p$ effective potential is obtained by interchanging $\rho_n\leftrightarrow\rho_p$. The $\Sigma^0 n\rightarrow\Sigma^0 n$ and $\Sigma^0 p\rightarrow\Sigma^0 p$ effective potentials have similar expressions, except that for their two-pion exchange parts the direct term as shown in fig.\,\ref{fig:3BFeff} vanishes and only an exchange diagram contributes in this sector.

The examples just discussed demonstrate the following. In the diagonal $\Lambda N\rightarrow\Lambda N$ potential the two low-energy constants, $H_1$ and $H_2$, those associated with the three-body contact term using decuplet dominance, appear in the combination $H_1 + 3H_2$. This leaves only one parameter in this channel, as it has been the case in previous calculations of $\Lambda$ single-particle potentials in nuclear and neutron matter performed in refs.\,\cite{Kohno2018,Haidenbauer2017}. In contrast,  the explicit inclusion of  $\Lambda NN\leftrightarrow\Sigma NN$ coupled channels introduces $H_1$ and $H_2$ in various different combinations so that these parameters have to be constrained independently.

\section{Brueckner-Hartree-Fock approach to hyperon single-particle potentials}

The two- and three-body potentials described in the previous section are used\footnote{We remind that due to direct and exchange contributions to the density-dependent effective $YN$ potentials, these have to be weighted with a statistical factor $1/2$ when added to the two-body potentials \cite{Kohno2018}.} as input for computing a hyperon-nucleon $G$-matrix within Brueckner theory at first order in the hole-line expansion (the Brueckner-Hartree-Fock approximation  \cite{Day1967}).
We focus on the single-particle potentials, i.e.\ the self-energies of hyperons in nuclear matter. The relevant formalism is briefly summarized below. For more details we refer to refs.\,\cite{Rijken1999,Schulze1998,Vidana2000,Kohno2000}.

The Brueckner reaction matrix or G-matrix is determined by solving the coupled-channel Bethe-Goldstone equation (in symbolic form)
\begin{equation}
G(\omega) = V + V \frac Q{e(\omega)+\mathrm i\epsilon} G(\omega)\,.
\end{equation}
The energy denominator \(e(\omega)\) depends on the starting energy \(\omega\).
The Pauli blocking operator \(Q\) excludes particles in intermediate states from scattering into the filled Fermi sea.
The potential $V$ is a matrix (labeled by the outgoing and ingoing two-baryon channels) including two-body and three-body contibutions, with the latter given in their density-dependent effective two-body form.

After angle-averaging (see appendix A for details) the Bethe-Goldstone equation decomposes into partial waves with total angular momentum \(J\): 
\begin{eqnarray}\label{eq:BGE}
&&G^{J}_{\alpha\beta}(p',p; P,\omega) = V^{J}_{\alpha\beta}(p',p)  +\\
&&\sum_{\nu}\int_0^\infty\frac{\mathrm d k\, k^2}{(2\pi)^3} \, V^{J}_{\alpha\nu}(p',k) 
\frac{\bar Q_{\nu}(P,k)}{\bar e_{\nu}(P,k;\omega)+\mathrm i \epsilon}G^{J}_{\nu\beta}(k,p;P,\omega) \,.\nonumber 
\end{eqnarray}
Here, $P$ is the total momentum of the two baryons. The indices $\alpha, \beta, \nu$ represent complete sets of channels, including partial wave quantum numbers and pairs of interacting baryons, $(B_1,B_2)$. In our case of interest, $B_1$ in the initial state is a $\Lambda$ hyperon, while $B_2$ is a nucleon within its Fermi sea. The channel coupling turns $B_1$ into a $\Sigma$ which subsequently interacts with the nuclear medium and turns back into a $\Lambda$ in the final state.

The $G$-matrix elements of eq.\,(\ref{eq:BGE}) are calculated at the on-shell starting energy
\begin{eqnarray}
\omega &=& E_{B_1}(p_1) +  E_{B_2}(p_2)~,\nonumber\\
\mathrm{with}~~~~~~E_{B_i}(p) &=& M_i + {p^2\over 2M_i} + \Re U_{B_i}(p)~,
\end{eqnarray}
where $M_i$ is the mass of baryon $B_i$.
Hence, the energy denominator $e(\omega)$ (see appendix A) requires the single-particle potentials for both hyperons and nucleons as an input. 

In Brueckner-Hartree-Fock approximation the single-particle potential for a baryon $B_1\in\{\Lambda,\Sigma^+,\Sigma^0,\Sigma^-,p,n\}$ interacting with the Fermi sea of nucleons $B_2\in\{p,n\}$ is then given by:
\begin{align}
&U_{B_1}(p_1) = 
\sum_{\alpha}\left[1+\delta_{B_1B_2}(-1)^{L+S}\right]\frac{(1+M_2/M_1)^3}{2}~\times\nonumber \\
& \sum_{J}(2J+1)
\int_{k_\mathrm{min}}^{k_\mathrm{max}}\! \frac{\mathrm dk\, k^2}{(2\pi)^3}\,W(p_1,k) \, G_{\alpha\alpha}^{J}(k,k;\bar P,\omega) \,.
\numberthis
\label{eq:U}
\end{align}
The weight function $W(p_1,k)$ resulting from the angular averaging procedure is specified in appendix A, together with the integration boundaries $k_\mathrm{min}$ and $k_\mathrm{max}$. Evidently, the calculation of the single-particle potentials depends on the single-particle potentials themselves as they appear in the energy denominator. Therefore eqs.~\eqref{eq:BGE} and \eqref{eq:U} must be solved self-consistently by iteration until convergence is reached. 
For the nucleon sector itself the $NN$ input potential is taken from ref.\,\cite{Entem2003}, employing SU(2) chiral perturbation theory at fourth order (N3LO). 
Chiral three-nucleon forces at order N2LO are also included. 

The computations are carried out using the so-called continuous choice, where single-particle potentials enter the energy denominator of intermediate states for all momenta, below and above the Fermi sea. From previous work \cite{Schulze1998} it is known that the continuous choice for intermediate states is preferable over the simpler gap choice (with potentials set to zero for momenta above the Fermi sea), because it allows for a reliable determination of the single-particle potentials including their imaginary parts. Some further technical details are given in Appendix B.

\section{Results}
\label{sec:res}
\subsection{Single-particle potentials of a $\Lambda$ hyperon in nuclear and neutron matter}
\label{subsec:spp}

\begin{figure*}[htpb]
\centering
\includegraphics[scale=.25]{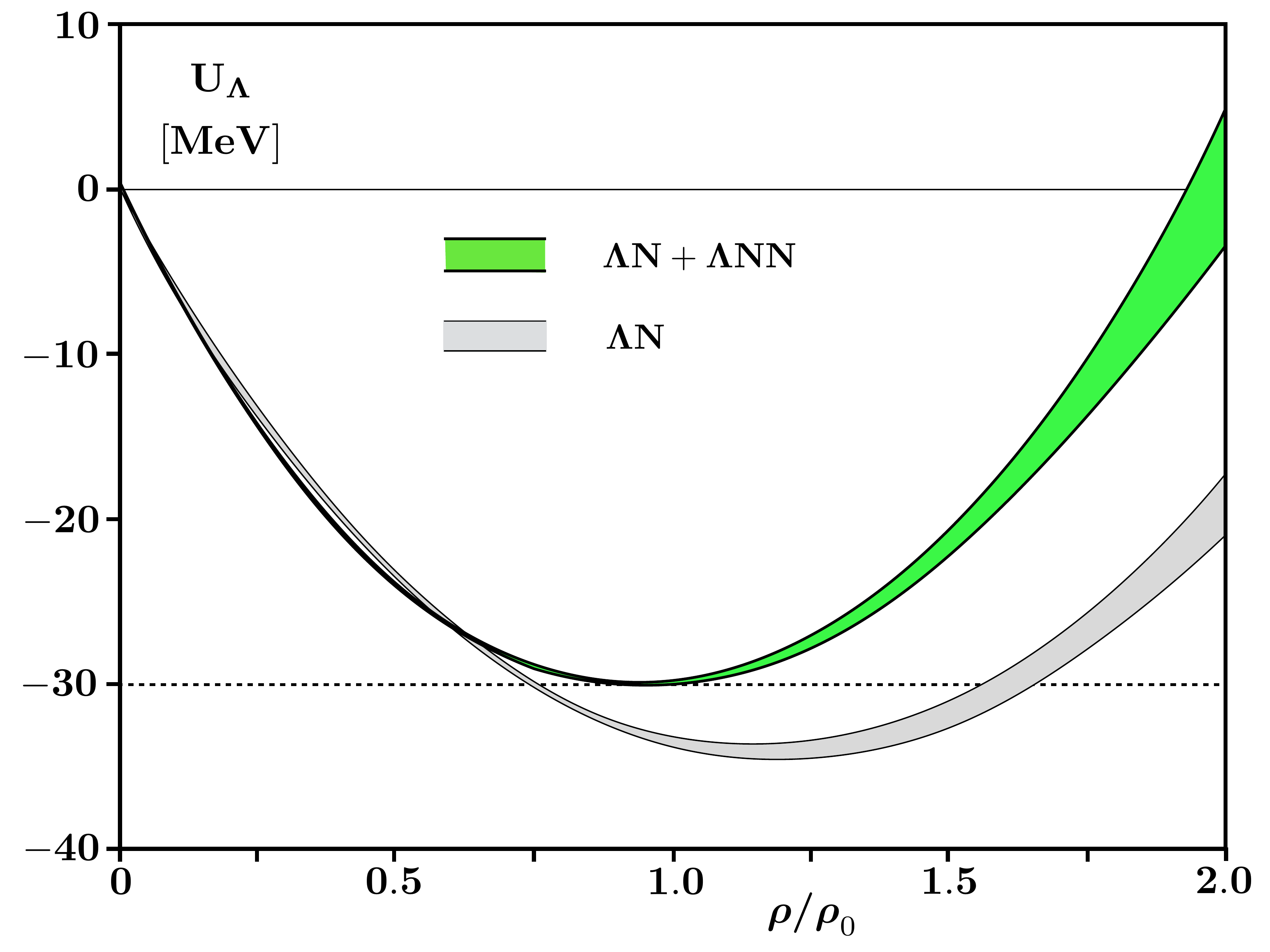} 
\caption{Single-particle potential $U_\Lambda(p=0;\rho)$ of a $\Lambda$ hyperon at rest in symmetric nuclear matter at densities up to $\rho = 2\rho_0 = 0.32$ fm$^{-3}$. Shown are results of the self-consistent solution of eqs.\,(\ref{eq:U}) and (\ref{eq:BGE}) using hyperon-nucleon two-body interactions NLO13 (lower band, $\Lambda N$), and in comparison with two- and three-body forces as input (upper band, $\Lambda N + \Lambda NN$). The calculations include explicitly the coupled channels $\Lambda N \leftrightarrow\Sigma N$ and $\Lambda NN\leftrightarrow\Sigma NN$. Upper and lower bands indicate uncertainties related to variations of cutoffs and choices of two representative pairs of low-energy constants, $(H_1,H_2) = (-1.2, 0)$ f$^{-2}$ and $ (-2.5, 1.2)$ f$^{-2}$, located on the consistency curves labelled NLO13 in Fig.\,\ref{fig:H1H2}. }
\label{fig:ULambda1}
\end{figure*}

The next step is now to calculate the single-particle potential, $U_\Lambda(p_1=0,\rho)$, for a $\Lambda$ at rest in symmetric nuclear matter as a function of baryon density $\rho=\rho_p+\rho_n$. Using as input the hyperon-nucleon potentials NLO13 and NLO19 together with the three-body interactions as specified in Section \ref{sec:int}, it is still necessary to constrain the two free parameters of the three-body contact terms. We recall that the calculation explicitly includes the $\Lambda NN \leftrightarrow\Sigma NN$ coupled channels, so the respective constants $H_1$ and $H_2$ enter independently. 

A key constraint comes from hypernuclear phenomenology. Bulk properties and the shell structure of $\Lambda$ hypernuclei are well described by a $\Lambda$-nuclear single-particle potential parametrized in Wood-Saxon form, with an accurately determined potential depth at nuclear central densities \cite{Gal2016}:
\begin{equation}
U_\Lambda(\rho \simeq \rho_0) = - 30~ \mathrm{MeV}~,
\label{eq:USNM}
\end{equation}
corresponding to about half of the attractive strength of the nucleon single-particle potential at nuclear matter saturation density, $\rho_0 = 0.16$ fm$^{-3}$. This constraint will be satisfied by any of the calculations to be described below.

The computations of $U_\Lambda$ based on self-consistent solutions of the Bethe-Goldstone equation (\ref{eq:BGE}) together with eq.\,(\ref{eq:U}) involve a momentum cutoff not only in the input hyperon-nucleon potentials but also in the momentum-dependent single-particle potentials $U(k)$ that appear in energy denominators and are integrated over intermediate momenta $k$. We apply a cutoff by a factor $\exp[-(k/\lambda')^6]$ and choose $\lambda'$ at 700 MeV, above the cutoff in the NLO hyperon-nucleon potentials, accomodating sufficient phase space for intermediate states in the Brueckner ladder while not altering the low-energy behavior. Variations around this cutoff, together with possible choices for the pair of constants ($H_1,H_2$), are reflected in the uncertainty estimates shown in subsequent figures.

Fig.\,\ref{fig:ULambda1} shows an instructive example of the calculated $U_\Lambda(p_1=0;\rho)$ in nuclear matter at low densities, using the NLO13 potential. Notably, the result using only the chiral two-body interaction would actually produce too much attraction in the $\Lambda$ potential, causing an overbinding of hypernuclei. This calls for repulsive effects beyond two-body interactions, with three-body forces being the natural extension. The three-body effects are moderate (a few MeV) at densities $\rho\simeq\rho_0$, but their relative importance grows continuously as the density increases. At $\rho \simeq 2\,\rho_0$ the repulsive $\Lambda NN$ 3BF is essential in order to turn the sign of $U_\Lambda$ from attractive to repulsive.

In the actual calculations the overbinding effect just mentioned turns out to be more pronounced for NLO19 than for NLO13. This implies that the repulsive three-body force needed to act in combination with NLO19 must be stronger than the one accompanying NLO13, in order to fulfill the hypernuclear constraint (\ref{eq:USNM}).  
The result in fig.\,\ref{fig:ULambda1} based on NLO13 is produced using particular examples of three-body constants, $(H_1,H_2)$, that evidently satify the constraint, eq.(\ref{eq:USNM}). However, the choice of $H_1$ and $H_2$ is not uniquely determined. It turns out that any pair $(H_1,H_2)$ that lies on either of the two curves labeled NLO13 in fig.\,\ref{fig:H1H2} is compatible with the condition, $U_\Lambda(\rho=\rho_0) = - 30$\,MeV. Furthermore, because of the difference in strengths required for the three-body forces that are associated with either the NLO13 or the NLO19 two-body interactions, the pairs of $(H_1,H_2)$ consistency curves for NLO19 and NLO13 are, not surprisingly, different. The question is then whether one can identify segments of these curves with parameter combinations $(H_1,H_2)$ such that the $\Lambda$ single-particle potential at higher densities becomes maximally repulsive. These are the candidate  parameters of choice for extrapolations to higher densities, as we are seeking mechanisms and conditions for avoiding a softening of the equation-of-state that would be in conflict with neutron star observations.

In practice the criteria for this scenario are set as follows. Here we focus on neutron matter. Consider the calculated $\Lambda$ single-particle potential at a sufficiently high density, $U_{\Lambda}(\rho = 3\,\rho_0)$, where the coupled-channel Bethe-Goldstone equation can still be solved reliably. A necessary condition for the non-occurrence of a $\Lambda$ in neutron matter at that density is that its chemical potential must satify $\mu_\Lambda(3\,\rho_0) = M_\Lambda + U_\Lambda(3\,\rho_0) > \mu_n( 3\,\rho_0)$. Anticipating the forthcoming analysis, it turns out that this required minimal condition implies $U_\Lambda(3\,\rho_0) > 65$ MeV. However, as this does not yet guarantee the suppression of $\Lambda$'s upon extrapolation to the higher densities as they are realized in the center of neutron stars, we set the condition more strictly as $U_\Lambda(3\,\rho_0) > 80$ MeV. It then turns out that high-density extrapolations of the $\Lambda$ chemical potential are indeed likely not to touch the neutron chemical potential. This stronger condition for maximally repulsive three-body forces, altogether consistent with hypernuclear phenomenology, is verified along the solid sectors of the curves in fig.\,\ref{fig:H1H2}. The dashed segments of these curves satisfy the weaker condition, $U_\Lambda(3\,\rho_0) > 65$ MeV, but may still encounter $\mu_\Lambda = \mu_n$ at some higher density.

The calculations of $U_\Lambda$ for both symmetric nuclear matter and pure neutron matter can be carried out safely (i.e. with well converging and numerically stable results) up to densities $\rho\simeq \rho_c=3.5\,\rho_0$. At even higher densities the occurrence of numerical instabilities limits the computations. One might argue that, in any case, performing calculations at $\rho\gtrsim 2\rho_0$ is not legitimate given the framework of chiral effective field theory with its limited range of applicability. On the other hand, the non-perturbative ladder summations of the Brueckner calculation lead beyond chiral perturbation theory, and the input interactions in the hyperon-nucleon sector are tested for total momenta $P$ up to and even above 800 MeV, whereas the neutron Fermi momentum even at central neutron star densities around $5\,\rho_0$ does not exceed 600 MeV, the typical cutoff scale in the calculations. We can therefore assume that at least qualitative extrapolations towards such high densities can be performed. 

\begin{figure*}[htpb]
\centering
\includegraphics[scale=.2]{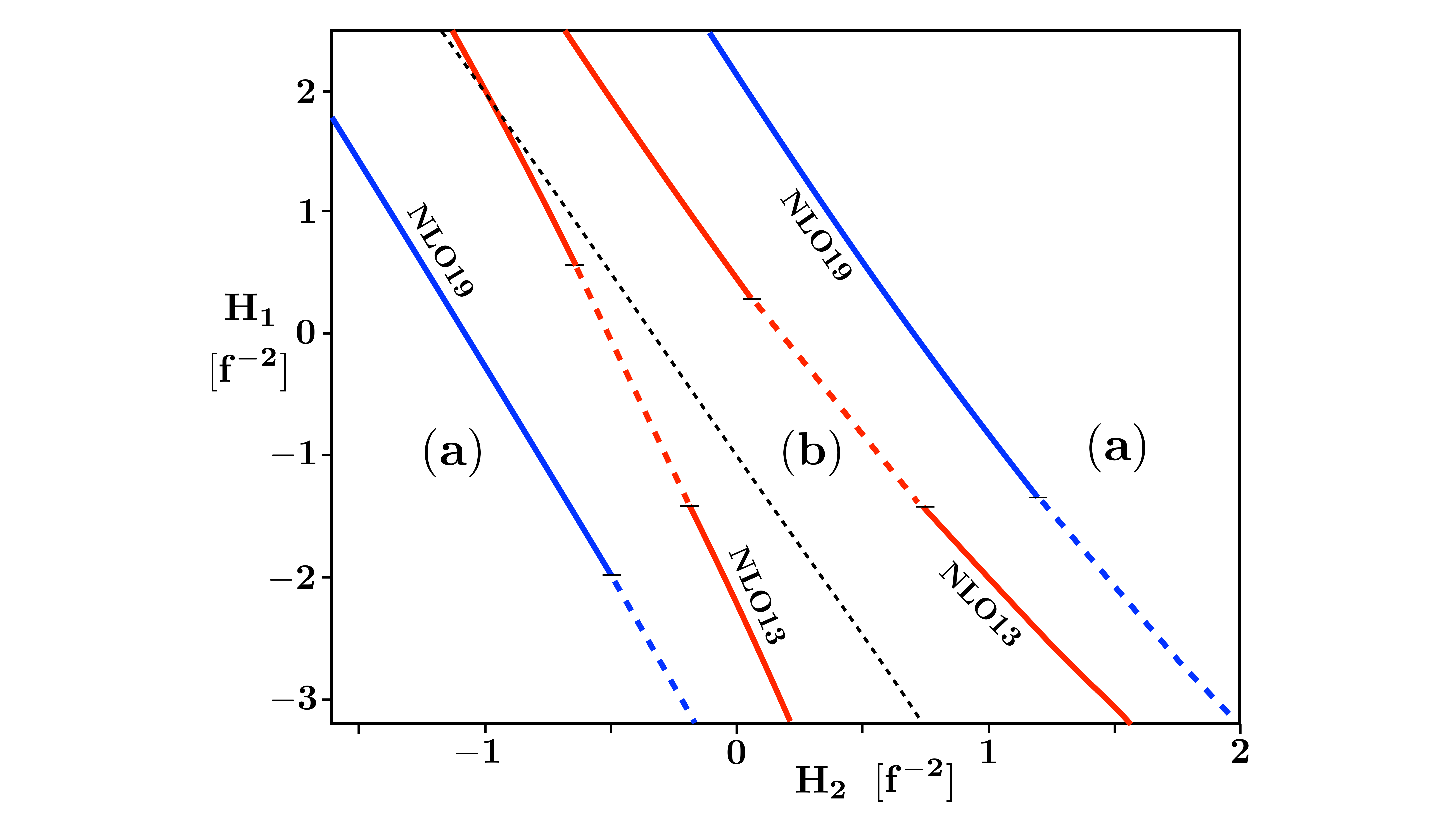} 
\caption{Plots of parameters $H_1$ and $H_2$ associated with the three-body contact terms discussed in section \ref{sec:int} (in units of $f^{-2} = 4.8$ fm$^2$, the inverse square of the pion decay constant). The two solid lines, each with labels NLO13 or NLO19, represent pairs $(H_1,H_2)$ that satisfy the constraint for the $\Lambda$ single-particle potential in symmetric nuclear matter, $U_\Lambda(\rho=\rho_0) = -30$ MeV, from hypernuclear phenomenology. The solid segments of the curves mark the combinations $(H_1,H_2)$ which generate maximally repulsive values for $U_\Lambda$ at high density, $\rho = 3\rho_0$, such that extrapolations to even higher densities maintain the condition $\mu_\Lambda>\mu_n$. In the dashed sections the three-body repulsion is not sufficiently strong  to suppress $\Lambda$'s in neutron stars.  Areas (a) and (b) indicate regions with $U_\Lambda(\rho_0)$ larger or less than $-30$\,MeV, respectively, to the left or right of the corresponding curves. The short-dashed line shows the linear dependence $H_1+3H_2= -f^{-2}$ \cite{Haidenbauer2017}, characteristic of the case in which only the diagonal $\Lambda NN$ three-body interaction is used (i.e. omitting $\Lambda NN\leftrightarrow\Sigma NN$ coupled channels).  }
\label{fig:H1H2}
\end{figure*}

For extrapolations to densities beyond the technically (numerically) accessible $\rho_c \simeq 3.5\,\rho_0$, we argue as follows. Around $\rho_c$, one finds clear evidence in both symmetric nuclear matter and neutron matter that the density dependence of $U_\Lambda$ has a prominent quadratic behaviour, $U_\Lambda(\rho) \propto \rho^2$. This is expected as the three-body terms begin to dominate at high densities and grow as $\rho^2$. The two-body terms at high density increase linearly with $\rho$, and this expected subleading behaviour is also observed in the numerical results already at densities beyond $2\,\rho_0$. Therefore, assuming continuity an extrapolation of $U_\Lambda$ to high denstites is supposed to be well justified using the following ansatz for $\rho > \rho_0$:
\begin{equation}
U_\Lambda(\rho) = u_0 + u_1\left({\rho\over\rho_0}-1\right) + u_2\left({\rho\over\rho_0}-1\right)^2\,. 
\label{eq:extrap}
\end{equation}
The parameters $u_1$ and $u_2$ are determined by fits to the calculated single-particle potentials at densities below $\rho_c$.
With $u_0$ fixed at $\rho=\rho_0$, it turns out that the high-density behavior is indeed governed by the quadratic term, with $u_2 \gg u_1$.

\begin{figure*}[htpb]
\centering
\includegraphics[scale=.25]{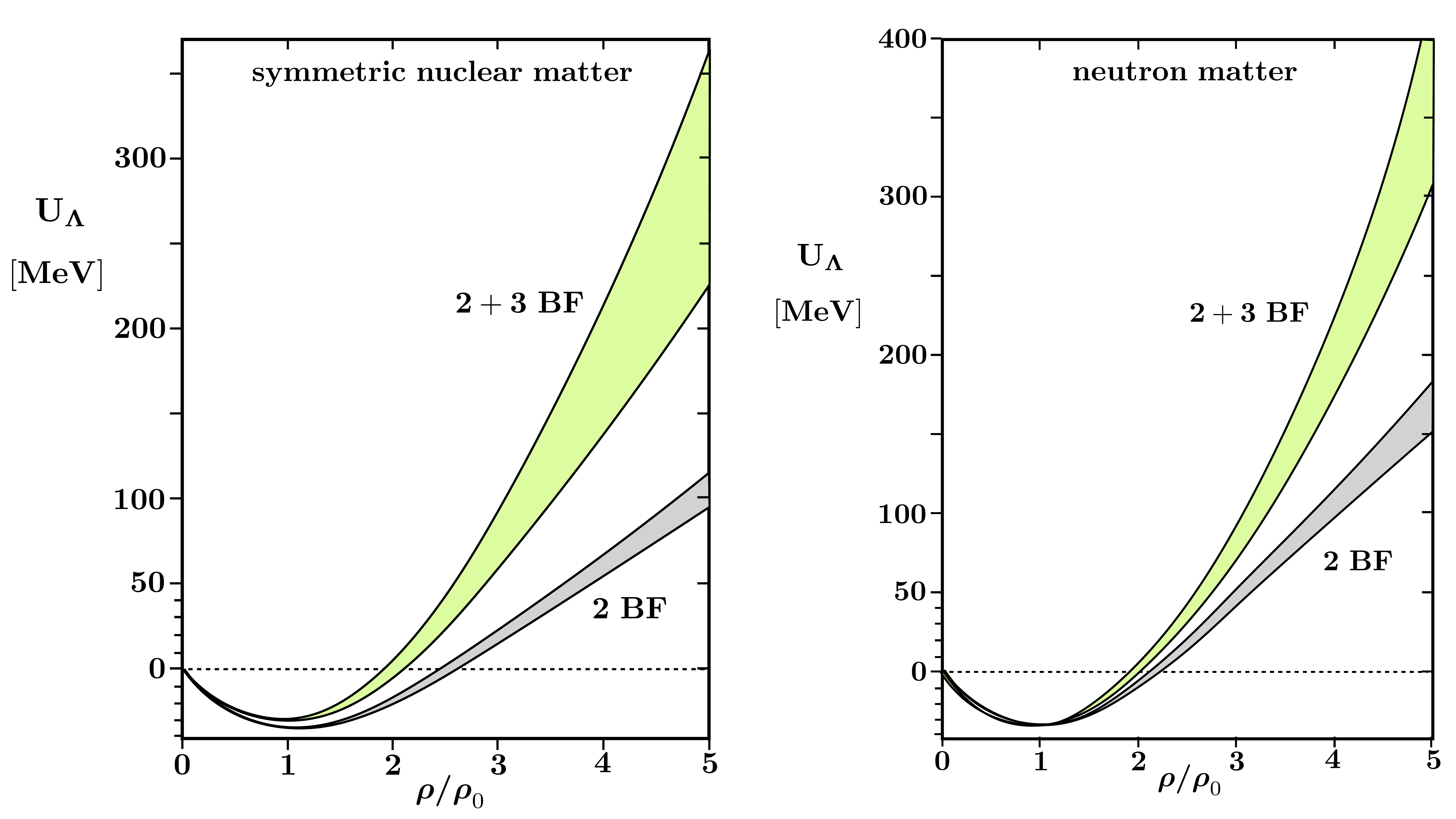} 
\caption{Single-particle potentials $U_\Lambda(p=0;\rho)$ of a $\Lambda$ hyperon in dense symmetric nuclear matter (left) and neutron matter(right), based on self-consistent solutions of eqs.\,(\ref{eq:U}) and (\ref{eq:BGE}) computed up to $\rho = 3.5\,\rho_0$ using the NLO13 interaction, and further extrapolated to higher densities as described in the text.
The uncertainty bands reflect cutoff dependence and choices of $(H_1,H_2)$  from the lower solid segments of the NLO13 lines of fig.\,\ref{fig:H1H2}.}
\label{fig:ulambda}
\end{figure*}

\begin{figure*}[htpb]
\centering
\includegraphics[scale=.25]{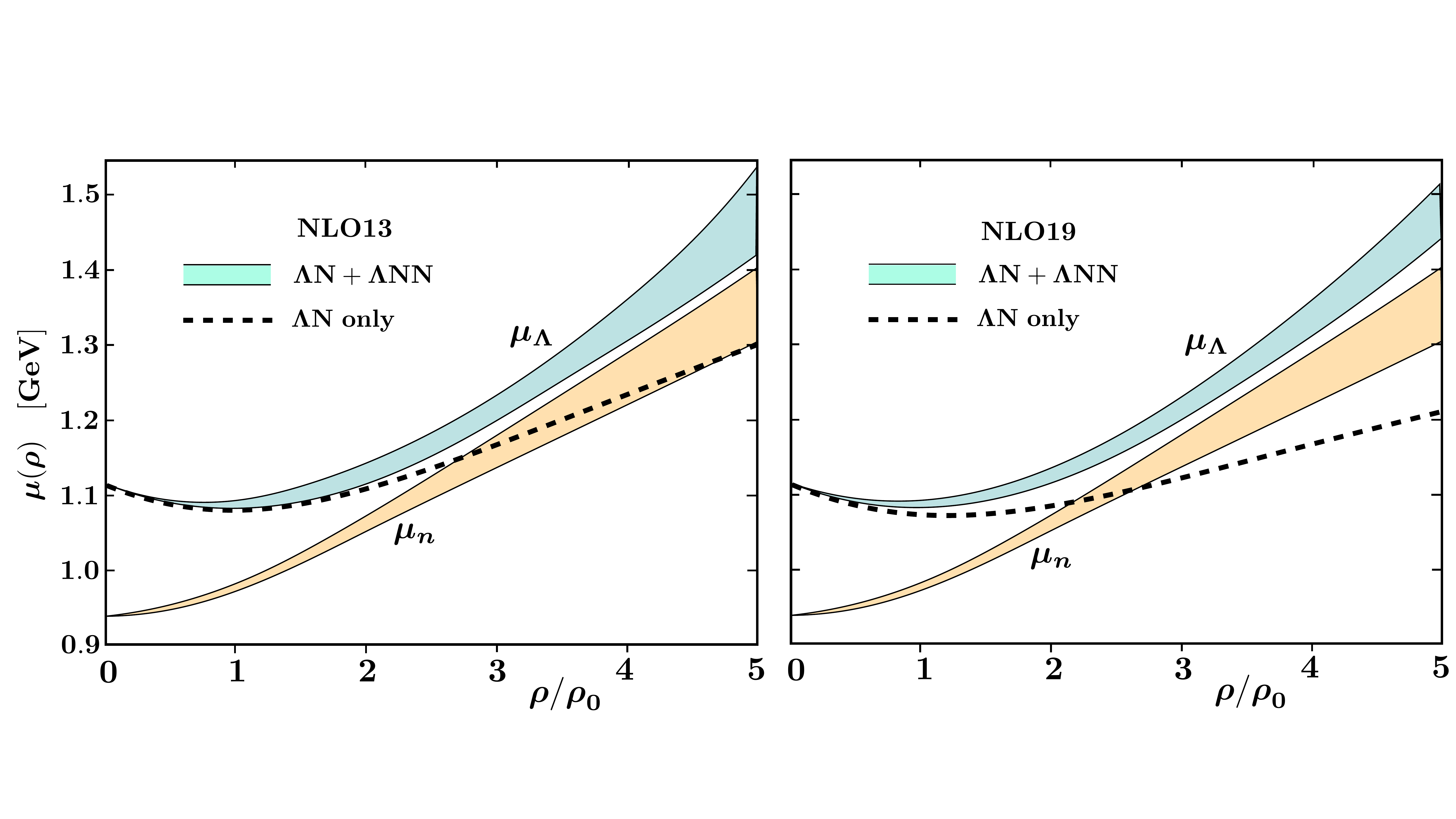} 
\caption{Comparison of $\Lambda$ and neutron chemical potentials, $\mu_\Lambda$ and $\mu_n$, in neutron star matter up to baryon densities typically encountered in the center of neutron stars. The neutron chemical potential is derived from the equation-of-state calculated in \cite{DW2017} using chiral SU(2) nucleon-meson field theory combined with functional renormalization group methods. The uncertainty band reflects primarily the errors in the nuclear symmetry energy, $E_S = 32\pm3$ MeV. The $\Lambda$ chemical potential is based on $U_\Lambda$ as in fig.\,\ref{fig:ulambda}, calculated using the chiral SU(3) interactions NLO13 (left panel) and NLO19 (right panel) with full two- and three-body forces ($\Lambda N +\Lambda NN$) and sets of three-body parameters as explained in the text. The dashed line shows $\mu_\Lambda$ refers to two-body $YN$ interactions only. }
\label{fig:chempot}
\end{figure*}

Results for the high-density extrapolations of $U_\Lambda$ in symmetric nuclear matter and neutron matter are displayed in fig.\,\ref{fig:ulambda}. With two-body $YN$ interactions only, the linearly increasing $U_\Lambda$ at densities $\rho\gtrsim 2\rho_0$ is reminiscent of a Hartree potential. When increasingly repulsive three-body interactions are added, the leading $\rho^2$ dependence of the 3BF takes over at high densities. Note that in neutron matter, the strong repulsion in $U_\Lambda$ exceeds 100 MeV already at $\rho \simeq 3.5 \,\rho_0$ and reaches $U_\Lambda > 300$ MeV at $\rho\simeq 5\,\rho_0$, the density range characteristic of the inner core of neutron stars.

\subsection{Hyperon and neutron chemical potentials in neutron stars}

If the chemical potentials of the $\Lambda$ hyperon and neutron fulfill the condition $\mu_\Lambda = \mu_n$ at some density in neutron star matter, it becomes energetically favourable to replace neutrons by $\Lambda$ hyperons via weak interactions. For a baryonic species $j$, the chemical potential $\mu_j$ is determined by the derivative of the energy density ${\cal E}$ with respect to the partial density $\rho_j$ of that species:
\begin{equation}
\mu_j = {\partial{\cal E}\over\partial\rho_j}~.
\end{equation}
Given a realistic neutron star equation-of-state, $P({\cal E})$, the Gibbs-Duhem relation for multicomponent sytems
\begin{equation}
P({\cal E}) + {\cal E} = \rho_n\,\mu_n + \rho_p\,\mu_p + \rho_\Lambda\,\mu_\Lambda\,+ \dots~~.
\end{equation}
determines the chemical potentials. In practice we use the microscopic EoS computed from a chiral nucleon-meson field theory in combination with functional renormalization group methods \cite{DW2017,FW2019}. This EoS is consistent with all important nuclear physics constraints and neutron star observations. For simplicity we neglect the small (few percent) proton fraction from beta equilibrium which is part of the neutron star EoS, but its effect is marginal in the present context. We shall further quantify this small correction later in this subsection.

At the possible onset of $\Lambda$ hyperons, their kinetic energy vanishes as there is no $\Lambda$ Fermi sea to start with. Therefore the minimal $\Lambda$ chemical potential at that point is simply given by
\begin{equation}
\mu_\Lambda(\rho) = M_\Lambda + U_\Lambda(\rho)~,
\label{eq:mulambda}
\end{equation}
where $M_\Lambda$ is the $\Lambda$ hyperon mass and $\rho\simeq\rho_n$ is the neutron density. The input $U_\Lambda(\rho)$ includes the one shown in fig.\,\ref{fig:ulambda} for neutron matter, but now uses both versions of chiral two-body $YN$ potentials, NLO13 and NLO19, and three-body forces with a wide range of parameters $(H_1,H_2)$ covering the solid segments of the curves displayed in fig.\,\ref{fig:H1H2}.

The coupled-channel Brueckner calculation of $U_\Lambda(\rho)$ based on NLO13 features a strong $\Lambda N\leftrightarrow\Sigma N$ coupling which is a major source of the $\Lambda N$ attraction at intermediate distances. Its partial suppression by the Pauli principle acting on intermediate nucleon states in nuclear or neutron matter is responsible for the turnover from attraction to repulsion in the $\Lambda$-nuclear two-body force at a density well below $3\,\rho_0$ in nuclear matter. From fig.\,\ref{fig:ULambda1} we also recall that the $\Lambda N$ attraction provided by NLO13 is slightly stronger than what is needed for the empirical hypernuclear shell-model potential. This sets the frame for constraining unknown pieces of the repulsive $YNN$ three-body force which in turn governs the behaviour of $\mu_\Lambda$ at high density. The alternative $YN$ potential NLO19 is equivalent to NLO13 with respect to $\Lambda N$ and $\Sigma N$ scattering data but produces a stronger attractive $\Lambda$-nuclear potential than NLO13 at $\rho\simeq\rho_0$. This difference is balanced by a more strongly repulsive three-body $\Lambda NN$ interaction correlated with NLO19. 

The comparison of $\mu_\Lambda$ and $\mu_n$ is shown in fig.\,\ref{fig:chempot}. The uncertainty band of the neutron chemical potential is related primarily to the range of possible values of the nuclear symmetry energy, $E_\text{{sym}} = (32\pm 3)$ MeV. We note that this uncertainty band also includes $\mu_n$ as given in Ref.\,\cite{APR1998} for their maximally repulsive interaction (AV18+$\delta v$ + UIX*) up to $\rho \lesssim 4\,\rho_0$.
 
Fig.\,\ref{fig:chempot} points out that the combined repulsion from two- and three-body hyperon-nuclear interactions for both NLO13 and NLO19 cases can indeed be potentially strong enough to avoid the appearance of $\Lambda$ hyperons in neutron stars. One finds $\mu_\Lambda > \mu_n$  throughout the neutron star density range when a set of three-body parameters is selected from the solid segments of the lines in Fig.\,\ref{fig:H1H2} that are constrained by hypernuclear spectroscopy. Two-body $\Lambda N$ interactions alone would not be sufficient to suppress the occurrence of $\Lambda$ hyperons. Their onset would appear already at densities around $3\,\rho_0$ {for NLO13} and even lower, at  $2\,\rho_0$, for NLO19. The resulting softening of the EoS would not be acceptable in comparison with observations of the heaviest neutron stars.

A further comment concerns $\Sigma$ hyperons in neutron star matter. At high density it is in principle possible to consider the condition
\begin{equation}
\mu_{\Sigma^-} = \mu_n+\mu_e = 2\mu_n -\mu_p
\end{equation}
for the occurrence of $\Sigma^-$ replacing an electron together with a neutron. However, this turns out not to be an option because the $\Sigma^-$ potential in neutron matter is strongly repulsive \cite{Petschauer2016b}, so that $\mu_{\Sigma^-}$ remains separated by at least 50 MeV from $\mu_n + \mu_e$ at all relevant densities.

We close this section with an assessment of assumptions and remaining uncertainties which so far set limitations on drawing more detailed quantitative conclusions.

a) The present analysis starts from a hyperon-nucleon interaction derived from chiral SU(3) effective field theory at next-to-leading order. Further extensions to NNLO and higher orders would require a substantial enlargement and improvement of the empirical hyperon-nucleon and hyperon-nuclear data base beyond its presently existing status. It is helpful being to work with two low-energy equivalent interactions, NLO13 and NLO19. Their different behavior at high densities gives an impression of related uncertainties which can be judged from their comparison in fig.\,\ref{fig:chempot}.

b) The $YNN$ three-body forces are so far not well constrained. The decuplet dominance approximation, though well motivated, relegates existing uncertainties to two parameters, $H_1$ and $H_2$. Hypernuclear phenomenology sets limited constraints as pointed out in fig.\,\ref{fig:H1H2}. More restrictive conditions from detailed analysis of hyperon-nuclear few-body systems would certainly be desirable at this point. What the present investigation underlines is that a certain range of parameter pairs $(H_1,H_2)$, those on the solid lines in fig.\,\ref{fig:H1H2}, possibly provide repulsive three-body interactions that are capable of supressing the occurrence of $\Lambda$ hyperons in neutron stars. This is presumably not the case for those parameter pairs located on the dashed line segments in fig.\,\ref{fig:H1H2}.

c) Extrapolations to high densities beyond $3\,\rho_0$ rely on the assumption of continuity (e.g. no phase transition) and take into account the expected leading density dependences of the hyperon chemical potential in the presence of two- and three-body forces. For each pair of applicable three-body parameters $(H_1,H_2)$ a different high-density extrapolation results. The uncertainty bands in fig.\,\ref{fig:chempot} incorporate this exploratory freedom.

d) The calculations of the $\Lambda$ chemical potential are carried out for the case of pure neutron matter. Applications to neutron stars would strictly require to include the proton fraction $x_p$ induced by beta equilibrium, $\mu_n = \mu_p + \mu_e$. In order to estimate these effects, we have been guided by $x_p$ of the APR EoS \cite{APR1998} which is particularly large, exceeding $x_p \simeq 0.1$ at densities $\rho > 3\,\rho_0$, and thus suitable to provide an upper limit of proton admixture effects (the proton fraction of the EoS of neutron star matter used in the present work is about half the $x_p$ of APR). For selected values of the three-body parameters, $(H_1,H_2) = (-2.2,0)\,f^{-2}$ and $(-2.5,1.2)\,f^{-2}$, one finds that the difference in the resulting $\Lambda$ single particle potentials between neutron matter and beta-stable matter ranges between about 1 MeV at $\rho = 2\,\rho_0$ and typically less than 10 MeV at $\rho = 3.5\,\rho_0$. These small variations fall within the uncertainty band of $\mu_\Lambda$ in fig.\,\ref{fig:chempot}.

e) Results of the computations shown in the preceding figures have primarily been obtained using cutoffs $\lambda =500$ MeV in the chiral YN interactions and $\lambda' = 700$ MeV for single-particle potentials in the Brueckner ladder. Variations of $\lambda$ up to 600 MeV imply changes within the given error bands. With upward variations of $\lambda'$ the repulsive strength of $U_\Lambda$ increases so that $\mu_\Lambda$ separates further from $\mu_n$ at high density.

\section{Summary and outlook} 
\label{sec:sum}

Investigations of the properties of hyperons in nuclear and neutron matter have been extended to high baryon densities with the quest of exploring the possible occurrence of hyperons in neutron stars. Previous calculations based entirely on two-body hyperon-nucleon interactions suggested an onset for hyperons already at relatively low densities (between two and three times $\rho_0$). As a consequence the resulting equation-of-state is significantly softened and cannot support two-solar-mass neutron stars. The present work addresses the question whether hyperon-nucleon three-body forces (3BF) can provide the necessary repulsion to stop $\Lambda$ hyperons from replacing neutrons, maintaining the required stiffness of the neutron star EoS.  

Calculations have been carried out within the self-consistent Brueckner-Hartree-Fock (BHF) approach using the continuous choice for intermediate spectra. The employed microscopic potentials for the hyperon-nucleon interaction are constructed from SU(3) chiral effective field theory at next-to-leading order. In addition, the pertinent N3LO nucleon-nucleon interaction is taken from chiral EFT. 

Our main focus lies on the role of three-body forces, their reduction to density-dependent effective two-body interactions and the explicit treatment of $\Lambda NN\leftrightarrow\Sigma NN$ coupled channels in the BHF matrix equations. Unknown 3BF parameters have been constrained by reproducing the depth, $U_\Lambda(\rho=\rho_0) \simeq -30$ MeV, of the $\Lambda$ single-particle potential in nuclear matter as deduced from the phenomenology of hypernuclei.

The detailed discussion of $\Lambda$ and neutron chemical potentials in neutron matter then indicates that the $\Lambda NN$ three-body forces can indeed develop sufficient repulsive strength to suppress the occurrence of $\Lambda$ hyperons in neutron stars. This is the case if the pertinent three-body parameters are selected within a certain range that can be quantified, complementary to a smaller parameter interval that would still permit the appearance of $\Lambda$'s in dense neutron star matter. A comparison using different versions of the chiral EFT-based $YN$ interactions (NLO13 versus NLO19) underlines these expectations. While we regard these findings as potentially promising steps towards a possible clarification of the so-called hyperon puzzle in neutron stars, we can nevertheless not yet claim that this problem has been "solved", given the approximations and uncertainties listed at the end of the previous section. Detailed properties of the hyperon-nuclear three-body interactions must still be better constrained by data, e.g. in hypernuclear few-body systems. However, one can begin to consider the steadily increasing number of observational facts about neutron stars as part of a developing empirical data base that sets progressively more restrictive conditions for mechanisms involving strangeness in dense baryonic matter.\\
\\

\appendix{\bf Appendix A: \\ Angular average in the Bethe-Goldstone equation}\\
\label{app:angularaverage}

Here, we present details of the angular averaging procedure that preceeds the derivation of eq.\,(\ref{eq:BGE}).
Total and relative momenta for two baryons $B_1$ and $B_2$ are introduced by
\begin{equation}
\vec P = \vec p_1 + \vec p_2\,,\quad
\vec k = \frac{\xi_{12}\vec p_1-\vec p_2}{1+\xi_{12}}\,,\quad
\xi_{12} = \frac{M_{2}}{M_{1}}\,.
\end{equation}

A standard approximation is applied to replace $Q/e$ by the ratio of angle-averaged quantities, $\bar Q/\bar e$.
The Pauli operator, averaged over the angle $\theta$ between $\vec P$ and $\vec k$, is given by
\begin{align*}
\bar Q_{\nu}(P,k) & =
\frac12\int_{-1}^1\!\mathrm d\cos\theta\
\Theta\big(|\vec p_1| - k_F^{(1)}\big)\Theta\big(|\vec p_2| - k_F^{(2)}\big) \\
 & = [0|\frac{[-1|z_1|1]+[-1|z_2|1]}2|1] \,,
\numberthis
\end{align*}
with 
\begin{align*}
z_1 & = \frac{1+\xi_{12}}{2kP}\bigg\{\left(\frac1{1+\xi_{12}}P\right)^2+k^2-(k_F^{(1)})^2\bigg\}\,, \\
z_2 & = \frac{1+1/\xi_{12}}{2kP} \bigg\{\left(\frac{\xi_{12}}{1+\xi_{12}}P\right)^2 + k^2-(k_F^{(2)})^2\bigg\} \,.
\numberthis
\end{align*}
and the notation \([a|b|c]\equiv\operatorname{max}(a,\operatorname{min}(b,c))\) introduced in ref.\,\cite{Schulze1998}.
The angle-averaged energy denominator is given by
\begin{align*}
\bar e_{\nu}(P,k;\omega) = \omega & - \frac{P^2}{2M_\nu}-\frac{k^2}{2\mu_\nu} -M_\nu \\
& - \Re U_{B_1}(\bar p_1)- \Re U_{B_2}(\bar p_2) \,,
\numberthis
\end{align*}
with \(M_{\nu}=M_1+M_2\) and \(\mu_{\nu}=M_1 M_2/(M_1+M_2)\). The angle-average is done for the arguments of the single particle potentials \(U_{B_i}\) of the intermediate baryons:
\begin{align*}
\bar p_1 
&= \left(\tfrac1{(1+\xi_{12})^2} P^2 + k^2 + 2\tfrac1{1+\xi_{12}} P k\, \overline{\cos\theta}\right)^{1/2} \,, \\
\bar p_2 
&= \left(\tfrac{\xi^2_{12}}{(1+\xi_{12})^2} P^2 + k^2 - 2\tfrac{\xi_{12}}{1+\xi_{12}} P k\, \overline{\cos\theta}\right)^{1/2} \,,
\numberthis
\end{align*}
with
\begin{align*}
\overline{\cos\theta} & 
= \frac{ \int_{-1}^1\!\mathrm d\cos\theta\, \cos\theta\, Q(\vec P,\vec k) }{ \int_{-1}^1\!\mathrm d\cos\theta\, Q(\vec P,\vec k) } \\
& = \frac12\big([-1|z_2|1]-[-1|z_1|1]\big) \,.
\numberthis
\end{align*}
where \(Q(\vec P,\vec k)\) is the exact Pauli blocking operator.

It is common practice to introduce a further simplification, replacing
the squared momenta $P^2=P^2(\vec p_1, \vec k)$ and $p^2_2=p^2_2(\vec p_1, \vec k)$ entering the Bethe-Goldstone equation by their angle averages:
\begin{align*}
\bar{P}^2(p_1,k) & = \frac{\int_{|\vec p_2|\leq k_F^{(2)}}\!\mathrm d\cos\vartheta\, P^2(p_1,k,\cos\vartheta)}{\int_{|\vec p_2|\leq k_F^{(2)}}\!\mathrm d\cos\vartheta} \\
&=
(1+\xi_{12})^2\left[p_1^2+k^2-p_1k(1+[-1|x_0|1])\right] \,, \\
\bar p_2^2(p_1,k)
&=
\frac{\xi_{12}}{1+\xi_{12}}\bar P^2(p_1,k) + (1+\xi_{12})k^2 -\xi_{12}\,p_1^2 \,,
\numberthis
\end{align*}
where \(\vartheta\) is the angle between \(\vec p_1\) and \(\vec k\), and one finds
\begin{equation*}
x_0 = \frac{\xi_{12}^2p_1^2+(1+\xi_{12})^2k^2-(k_F^{(2)})^2}{2\xi_{12}(1+\xi_{12})p_1k} \,.
\numberthis
\end{equation*}
Note that baryon \(B_2\) in the initial state is within its Fermi sea.

The weight function $W(p,k)$ that appears in the calculation of the single-particle potential $U_B(p)$ via eq.\,(\ref{eq:U}) is given by
\begin{align*}
W(p,k) &= \frac1{4\pi}\int_{|\vec p_2|\leq k_F^{(2)}}\!\mathrm d\Omega_k = \frac12(1-[-1|x_0|1]) \,.
\numberthis
\end{align*}
The integration boundaries, \(k_\mathrm{min}\) and \(k_\mathrm{max}\),  of the relative momentum are determined by the condition \(W(k_1,k) = 0\), which leads to
\begin{align*}
k_\mathrm{min} &= \operatorname{max}\left( 0, \frac{-k_F^{(2)}+\xi_{12} p}{1+\xi_{12}} \right) \,,\
k_\mathrm{max} = \frac{k_F^{(2)}+\xi_{12} p}{1+\xi_{12}} \,.
\numberthis
\end{align*}

\appendix{\bf Appendix B: Some technical details}\\

Eqs. (\ref{eq:BGE},\ref{eq:U}) are solved numerically by alternately iterating both equations until the potentials $U_B$ converge. For certain kinematical conditions, e.g. involving large momenta in the initial state, the energy denominator in eq.\,(\ref{eq:BGE}) vanishes, giving rise to a pole in the Bethe-Gold- stone equation.
To make the integral numerically manageable, the principal value prescription is utilized:
\begin{equation}
	-\!\!\!\!\!\!\int_0^\infty \!dk \frac{N(k)}{D(k)} = 
	\int_0^\infty \!dk \left(\frac{N(k)}{D(k)}-\frac{N(k_0)}{D^\prime(k_0)} \frac{2k_0}{k^2-k_0^2}\right)\,, \end{equation}
for a simple pole at $k_0$, i.e. $D(k_0)=0$ and $D'(k_0)\neq0$.
This prescription eventually meets its limits at densities beyond $3.5\,\rho_0$, once second-order poles begin to show up.

For densities $\rho > 2\rho_0$ in pure neutron matter, slow convergence of $U_B$ occurs
because of increasingly strong input potentials. Convergence is improved by averaging $U_B(k)$ over subsequent iterations. High-density calculations occasionally require intervention by estimating an appropriate starting point for iterations from lower-density results. However, numerical stability is generally not guaranteed any more for $\rho > \rho_c \simeq 3.5\,\rho_0$. For densities $\rho > \rho_c$ we use the power series extrapolation in eq.(\ref{eq:extrap}), fitted to numerically stable results at $\rho \lesssim \rho_c$ as described in the text.

As discussed in ref. \cite{Kohno2018}, the cutoff $\lambda$ in the chiral input potential causes numerical oscillations of $U_B(k)$ for momenta $k > \lambda$. These oscillations tend to slow down the convergence of $U_B$ and are of no physical relevance. They are suppressed by the additional cutoff factor $\exp[-(k/\lambda_{spp})^6]$ attached to the single-particle potentials as described in Section \ref{subsec:spp}.\\

\acknowledgement{{\bf Acknowledgements}\\ We thank Johann Haidenbauer for helpful and instructive communications. This work is supported in part by DFG and NSFC (CRC110), and the DFG Excellence Cluster ORIGINS.}

\end{document}